# Constructing Gazetteers from Volunteered Big Geo-Data based on Hadoop

Song Gao [a], Linna Li [b], Wenwen Li [c], Krzysztof Janowicz [a], Yue Zhang [c]

[a] STKO Lab, Department of Geography, University of California, Santa Barbara, CA, USA

[b] California State University, Long Beach, CA, USA

[c] GeoDa Center for Geospatial Analysis and Computation, School of Geographical Sciences and Urban Planning, Arizona State University, Tempe, AZ, USA

**Abstract:**

[1] Traditional gazetteers are built and maintained by authoritative mapping agencies. In the age of Big Data, it is possible to construct gazetteers in a data-driven approach by mining rich volunteered geographic information (VGI) from the Web. In this research, we build a scalable distributed platform and a high-performance geoprocessing workflow based on the Hadoop ecosystem to harvest crowd-sourced gazetteer entries. Using experiments based on geotagged datasets in Flickr, we find that the MapReduce-based workflow running on the spatially enabled Hadoop cluster can reduce the processing time compared with traditional desktop-based operations by an order of magnitude. We demonstrate how to use such a novel spatial-computing infrastructure to facilitate gazetteer research. In addition, we introduce a provenance-based trust model for quality assurance. This work offers new insights on enriching future gazetteers with the use of Hadoop clusters, and makes contributions in connecting GIS to the cloud computing environment for the next frontier of Big Geo-Data analytics.

---

[1] The preprint version may differ from the final version for a peer-reviewed journal.

Corresponding author: Song Gao, e-mail: sgao@geog.ucsb.edu

# 1. Introduction

Place is a fundamental concept in daily life and reflects the way humans perceive, experience and understand their environment (Tuan, 1977). Place names are pervasive in human discourse, documents, and social media when location needs to be specified and referred to. Digital gazetteers are dictionaries of georeferenced place names, and play an important role in geographic information retrieval (GIR), in digital library services, and in systems for spatio-temporal knowledge organization (Hill, 2006; Goodchild & Hill, 2008; Li, Yang, & Zhou, 2008; Li, Raskin, & Goodchild; 2012). Several well-known authoritative digital gazetteers have been developed such as the Alexandria digital library (ADL) gazetteer at the University of California Santa Barbara (Hill, Frew, & Zheng, 1999; Goodchild, 2004), the Getty Thesaurus of Geographical Names (TGN) at the Getty Research Institute, the gazetteer maintained by the US Board on Geographic Names (BGN), and a Chinese gazetteer, KIDGS, at Peking University (Liu et al., 2009b). Such authoritative projects require expert teams to make lengthy efforts and the maintenance costs are high, thus often leading to lengthy delays in updating the databases.

With the emergence of the *social Web*, new forms of crowd-sourced gazetteers have become possible. They can be categorized in two types. One is collaborative mapping platforms, such as Wikimapia [2] and OpenStreetMap (OSM) [3], in which volunteers create and contribute geographic features and detailed descriptions to websites where the entries are synthesized into databases. The other way is socially constructed place, that is, gazetteer entries constructed from the Web documents and diverse social-media sources (such as Facebook, Twitter, Foursquare, Yelp, and Flickr) where the general public uses place names, describes sense of place, and makes diverse comments according to their experiences (Uryupina, 2003; Jones et al., 2008; Goldberg, Wilson, & Knoblock, 2009; Li, Goodchild, & Xu, 2013). Note that the term *gazetteer* in this paper also includes *point of interest* (POI) databases such that the **P** stands for place not point. By mining such rich resources, it is possible to construct or enrich gazetteers in a bottom-up approach instead of in a traditional top-down approach (Adams & Janowicz, 2012; Adams & McKenzie, 2013). However, the data mining and harvesting processes are computationally intensive. Especially in the age of Big Data, the volume, the updating velocity, and the variety of data are too big, too fast and too (semantically and syntactically) diverse for existing tools to process (Madden, 2012). In the GIScience/GIS community, researchers may not be willing to wait for weeks or longer to process the terabyte or petabyte-scale geotagged data streams. Fortunately the emerging cloud-computing technologies offer scalable solutions for some of the processing problems in Big Data Analytics.

---

[2] http://www.wikimapia.org

[3] http://www.openstreetmap.org

In this research, we present a novel approach to harvest crowd-sourced gazetteer entries from social media and to conduct high-performance spatial analysis in a cloud-computing environment. The main contribution of this paper is two-folds: First, it introduces the design and implementation of a scalable distributed-platform based on Hadoop for processing Big Geo-Data and facilitating the development of crowd-sourced gazetteers. Second, it provides valuable demonstrations about how to efficiently extract multiple feature types of gazetteer entries at multiple scales and how to integrate emerging data and technologies to improve GIScience research.

The rest of the paper is organized as follows. In Section 2, we introduce some relevant work about space and place, gazetteers, VGI, and Big Data, as well as cloud-computing infrastructures, to help understand the challenges involved in the presented research. In Section 3, we design and implement a novel Hadoop-based geoprocessing platform for mining, storing, analyzing, and visualizing crowd-sourced gazetteer entries; this is followed by experiments and results, as well as a trust evaluation in Section 4. We conclude the paper with discussions and directions for future research (Section 5).

## 2. Related work

In this section we briefly point to related work and background material.

*2.1 Space and place*

Space and place are two fundamental concepts in geography, and more broadly in the social sciences, the humanities, and information science (Tuan, 1977; Harrison, & Dourish, 1996; Goodchild & Janelle, 2004; Hubbard, Kitchin, & Valentine, 2004; Agnew, 2011; Goodchild, 2011). The spatial perspective is studied based on geometric reference systems that include coordinates, distances, topology, and directions; while the alternative "platial" (based on place) perspective is usually defined by textual place names, linguistic descriptions, and the semantic relationships between places (Janowicz, 2009; Goodchild and Li, 2012a; Gao et al., 2013). There would not be any places without people's perception and cognition. As argued by Tuan (1977), it is humans' interactions and experiences that turn space into place. Place is not just a thing in the world but a social and cultural way of understanding the world. Giving names and descriptions to locations is a process to make space meaningful as place. Social-tagging, tweets, photo sharing, and geo-social check-in behaviors have created a large volume of place descriptions on the Web.

Researchers have made significant efforts toward georeferencing place descriptions and processing spatial queries, such as using ontologies of place (Jones, Alani, & Tudhope, 2001), using a qualitative spatial reasoning framework (Yao & Thill, 2006), using fuzzy objects (Montello et al., 2003), using probability models in combination with uncertainty (Guo, Liu, & Wieczorek, 2008; Liu et al., 2009a), using kernel-density estimation (Jones et al., 2008), using description logics (Bernad et al., 2013), as well as knowledge discovery from data techniques for platial search (Adams & McKenzie, 2012). Recently, a review by Vasardani, Winter, and Richter (2013) has suggested that a synthesis approach would provide improvements in locating place descriptions, and that new opportunities exist in identifying places from public media and volunteered sources by using Web-harvesting techniques.

*2.2 Gazetteers*

Existing GIS and spatial databases are mature in representing space, but limited in representing place. In order to locate place names on a map with precise coordinates and to support GIR, efforts have been taken to convert place to space. One major mechanism is the use of gazetteers, which conventionally contain three core elements: place names (N), feature types (T), and footprints (F) (Hill, 2000). A place name is what people search for if they intend to learn about a place, especially its location, in a gazetteer. A place type is a category picked from a feature-type thesaurus for classifying similar places into groups according to explicit or implicit criteria. Janowicz and Keßler (2008) argued that an ontological approach to defining type classifications will better support gazetteer services, semantic interoperability (Harvey et al., 1999; Scheider, 2012), and semi-automated feature annotation. A footprint is the location of a place, and is almost always stored as a single point which represents an extended object as an estimated center, or the mouth in the case of a river. Recent work is providing additional spatial footprints including polygons and part-of relations.

One major role of a gazetteer is thus to link place names to location coordinates. For example, the ADL model which links places to spatially defined digital library resources requires a comprehensive gazetteer as part of its spatial query function to provide access to web services, including collections of georeferenced photographs, reports relating to specific areas, news and stories about places, remote sensing images, or even music (Goodchild, 2004). The minimum required elements of a place in ADL model are represented by the triples <N, T, F>. As a start, ADL combines two databases: the Geographic Names Information System (GNIS) and the Geographic Names Processing System (GNPS), both from US federal-government agencies. Frequently, it is necessary to consult and combine results from multiple gazetteer sources, which is generally described as (feature) conflation (Saalfeld, 1988). Hastings (2008) has proposed a computational framework for automated conflation of digital gazetteers based on three types of similarity metrics: geospatial, geotaxial, and geonomial. In addition, efforts have been made in mining gazetteers semi-automatically from the Web (e.g., Uryupina, 2003; Goldberg, Wilson, & Knoblock, 2009). Challenges such as interoperability and quality control need to be investigated in such crowd-sourced gazetteers. The conflation of POI databases is widely considered an important next research step to combine the different attributes stored by various systems to more powerful joint database.

*2.3 Big Data and VGI*

Big Data is used to describe the phenomenon that large volumes of data (including structured, semi-structured, and unstructured data) on various aspects of the environment and society are being created by millions of people constantly, in a variety of formats such as maps, blogs, videos, audios, and photos. Big Data is "big" not only because it involves a huge amount of data, but also because of the high dimensionality and inter-linkage of a multitude of (small) datasets that cover multiple perspectives, topics, and scales (Janowicz et al., 2012). The Web has lowered previous barriers to the production, sharing, and retrieval of varied information linked to places. VGI (Goodchild, 2007), a type of user-generated content (UGC) with a geospatial component, has gradually been taking the lead as the most voluminous source of geographic data. For example, there were over 20 million geographic features in the database of Wikimapia at the time of writing, which is more than many of the world's largest gazetteers. In addition to features with explicit locational information stored in geodatabases, places are also mentioned and discussed in social media, blogs, and news forums, etc., but many of the places referenced in this way do not appear in official gazetteers. This type of unstructured geographic information is rich and abundant, with a great potential to benefit scientific research and decision making.

This phenomenon provides a great potential to advance research on gazetteers. Although gazetteers provide a convenient way to link place names and locations, there are limitations in official place descriptions. The intended use of an authoritative gazetteer is to facilitate communication between government agencies, so only clearly defined geographic features that are important for policy making are included, e.g. administrative divisions and boundaries. Some places that are commonly referred to in daily conversations may not be considered (e.g., coffee shops). In addition, new place names emerging from popular cultures cannot be added to an official gazetteer in a timely manner because it is time-consuming to make changes by holding board meetings to discuss adjustments. Another missing function of official gazetteers is the representation of vague spatial extents of places. Fortunately, the limitations of official gazetteers might be partially complemented by integrating new sources based on VGI. For example, Keßler, Janowicz, and Bishr (2009) have proposed an agenda for an infrastructure of next-generation gazetteers which allow bottom-up contributions by incorporating volunteered data.

*2.4 Cloud computing and CyberGIS*

Cloud computing services and their distributed deployment models offer scalable computing paradigms to enable Big Data processing for scientific researches and applications (Armbrust et al., 2010; Ostermann et al., 2010), thus offering opportunities to advance gazetteer research. Some representative cloud systems and the characteristics of clusters, grids, cloud systems have been carefully examined by Buyya et al. (2009). Cloud services can be categorized into three main types: infrastructure as a service (IaaS), platform as a service (PaaS) and software as a service (SaaS). IaaS, as used in this work, provides the access to computing hardware, storage, network components and operating systems through a configurable virtual server. An IaaS user can operate the virtual server, install software tools, configure firewalls, and run model simulations remotely as easily as accessing a physical server. More importantly, it is more convenient for researchers to utilize these scalable cloud-computing resources with the availability of low-cost, on-demand IaaS such as the Web services of the Amazon elastic computing cloud (AWS EC2) and Amazon simple storage service (Amazon S3).

In the geospatial research area, cloud computing has attracted increasing attention as a way of solving data-intensive, computing-intensive, and access-intensive geospatial problems (Yang et al., 2011a). For example, in order to enhance the performance of a gazetteer service, Gao et al. (2010) designed a resource-oriented architecture in a cloud-computing environment to handle multiple levels of place-name queries. Yang et al. (2011b) presented how spatial computing facilitates fundamental physical science studies with high-performance computing capabilities. The emerging concept of *CyberGIS*, which synthesizes cyberinfrastructure, spatial analysis, and high-performance computing, provides a promising solution to aforementioned geospatial problems as a cloud service (Yang et al., 2010; Wang, 2010; Li et al., 2013). Scalable and efficient geo-processing is conducted on the high-end computing facilities and released as standard Web services; a Web portal is provided to Internet users to interact with the servers, upload/download raw data, perform analysis, and visualize results. From this perspective, the CyberGIS gateway can be considered a combination of IaaS, PaaS, and SaaS and its architecture provides guidance for establishing other cloud geoprocessing platforms. Several works conducted on the CyberGIS platform for Big Geo-Data analysis are presented in literature. For instance, Rey et al. (2013) discussed the parallelization of spatial analysis library—PySAL in multiple-core platforms. Liu and Wang (2013) described the implementation of a scalable genetic algorithm in HPC clusters for political redistricting. Wang et al. (2013) reviewed several key CyberGIS software and tools regarding to the integration roadmap.

There are many Big Data analytics platforms and database systems emerging in the new era, such as *Teradata* data warehousing platform, *MongoDB* No-SQL database, *IBM InfoSphere, HP Vertica, Red Hat ClusterFS* and *Apache Hadoop-based systems* like *Cloudera* and *Splunk Hunk*. They can be classified into two categories: (1) the massively parallel processing data warehousing systems like *Teradata* are designed for holding large-scale structured data and support SQL queries; and (2) the distributed file systems like *Apache Hadoop*. The advantages of Hadoop-based systems mainly lie in its high flexibility, scalability, low-cost, and reliability for managing and efficiently processing a large volume of structured and unstructured datasets, as well as providing job schedules for balancing data, resource and task loads. A MapReduce paradigm (more details in Section 3) implemented on Hadoop helps shift processing jobs to other connected nodes if one fails, such that it is inherently fault-tolerance. Compared with parallel relational-database-management-systems (DBMS) which perform excellently in executing a variety of data-intensive query processing benchmark (Pavlo et al., 2009), the Hadoop ecosystem is more optimized for computationally intensive operations such as geometric computations (Aji et al., 2013). However, such platforms have not been utilized thoroughly to process crowd-sourced Big Geo-Data, and little research has been conducted to construct gazetteers using such advanced cloud-computing platforms. In this research, we present how to build a scalable platform in detail to harvest and analyze crowd-sourced gazetteer entries based on the geoprocessing-enabled Hadoop ecosystem (GPHadoop).

## 3. The Hadoop-based processing platform

In this section we discuss the role and setup of Hadoop for the presented research.

*3.1 System architecture*

The goal of this processing platform is to provide a scalable, reliable, and distributed environment for mining, storing, analyzing, and visualizing gazetteer entries extracted from various Web resources (e.g., semi-structured geotagged data or unstructured documents). The system should also have the capability of processing geospatial data and an easy-to-use, configurable user interface to submit processing jobs and to monitor the status of the system. The open-source Hadoop is an ideal choice, since it provides a distributed file system and a scalable computation framework by partitioning computation processes across many host servers which are not necessary high-performance computers (White, 2012). More importantly, the *move-code-to-data* philosophy which applies within the Hadoop ecosystem will improve the efficiency since it usually takes more time to move voluminous data across a network than to apply the computation code to them. However, raw Hadoop-based systems usually lack powerful statistics and visualization tools (Madden, 2012). Therefore, we cannot use the raw Hadoop Cluster directly for Big Geo-Data analytics. Alternatively, we integrate the recently released Esri Geometry APIs [4] to spatially enable the Hadoop cluster for scalable processing of geotagged data from VGI sites and automatically link the results to an ArcGIS Desktop for visualization.

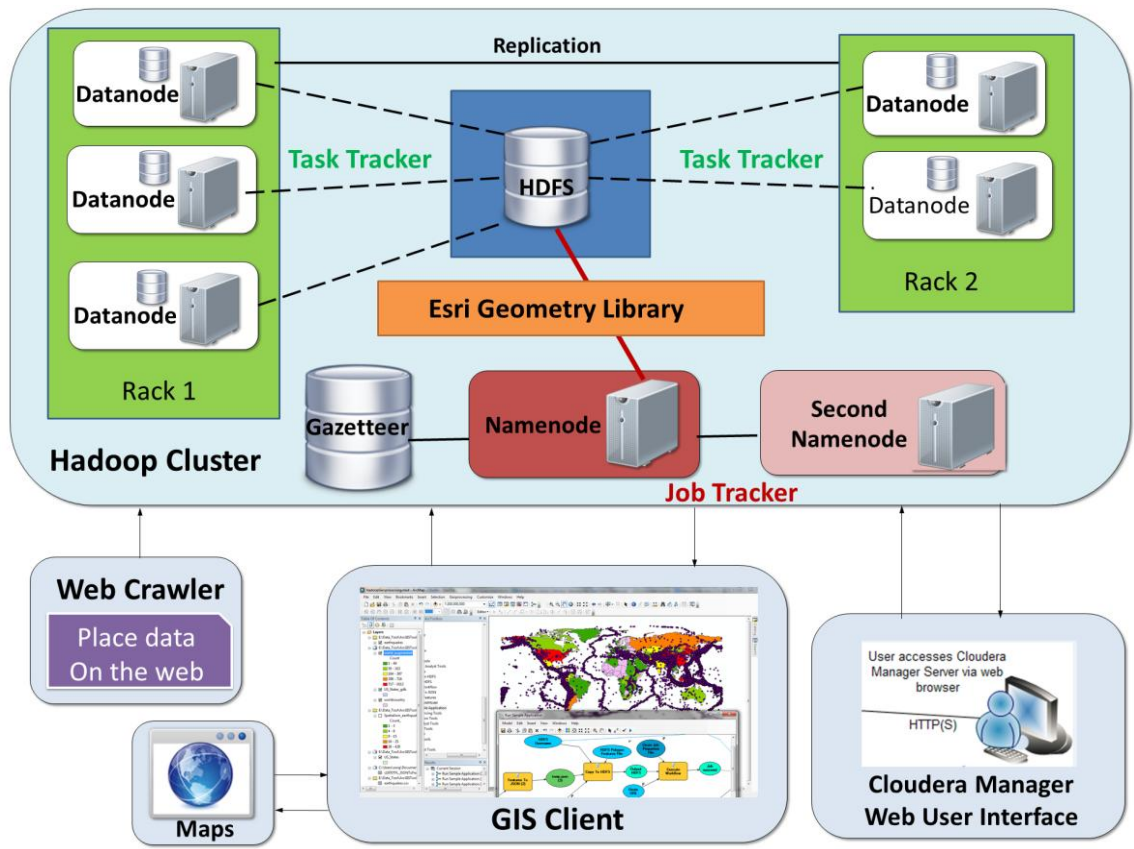

**Fig. 1.** System architecture

Fig. 1 demonstrates the system architecture of our Hadoop-based distributed geoprocessing platform (GPHadoop). It is composed of four modules: a Web crawler, a Hadoop cluster, a user interface supported by Cloudera and a GIS client.

(1) The Web crawler is a search engine written in Python to download place data from the Web and store them on the server. The Web crawler can process two types of data streams: unstructured textual place descriptions from Web documents or semi-structured data extracted from social media, e.g., Twitter's geotagged tweets and Flickr's geotagged photos[5]. Note that pre-processing and filtering (such as removing invalid coordinates) is necessary.

---
[4] https://github.com/Esri/geometry-api-java

(2) The Hadoop Cluster is the corpus of all server nodes within a group (their physical locations can differ) on Hadoop. Two Hadoop components -- the Hadoop distributed file system (HDFS) and the MapReduce programming model -- are implemented on our platform. HDFS is a distributed storage system for reliably storing and streaming petabytes of both unstructured and structured data on clusters (Shvachko et al., 2010). HDFS has three classes of nodes in each cluster:

- Name node: responsible for managing the whole HDFS metadata like permissions, modification and access times, namespace and disk space quotas. The most important role is to support the Web-HDFS access from the client via the cluster's public hostname, e.g. namenode.geog.ucsb.edu.
- Secondary name node: responsible for checking the name node's persistent status and periodically downloading current name-node image and log files; it cannot play the role of the primary name node.
- Data nodes: responsible for storing the unstructured file data or other structured data such as spreadsheets, XML files, and tab-separated-value files (TSV) in which the geotagged datasets have been stored. HDFS stores these files as a series of blocks (the unit of storage), each of which is by default 64MB (or 128MB) in size.

The MapReduce programming model is implemented on our platform for simplified processing of large Web datasets with a parallel, distributed algorithm on the Hadoop cluster (Dean & Ghemawat, 2008). Using MapReduce, a processing task is decomposed into *map*[6] and *reduce* sub-processes. In the *map* procedure, the name-node server divides the input into smaller sub-problems by generating intermediate key/value pairs and distributes them to data-nodes for solving sub-problems, while the *reduce* procedure merges all intermediate values associated with the same key, and passes the answer back to its master name node.

---

[5] http://www.flickr.com/services/api

In crowd-sourced gazetteers, processing text-based place descriptions is a computation-intensive procedure. For example, in order to identify how people are most likely to describe the characteristics of a place (e.g., the city of Paris), we need to calculate and rank the co-occurrence of tags that include the keyword of place name (e.g. Paris) across multiple documents. The MapReduce model can help to speed up this process. In the Algorithm 1, the *Mapper* function distributes the task of looping all the documents for calculating the co-occurrence frequency of words over multiple nodes and then the *Reducer* function will combine the results from all distributed nodes when they finish the parallel calculation. By using this algorithm, the most popular words to describe a place can be identified very quickly.

**Algorithm 1:** the MapReduce algorithm for co-occurrence word counting.

---

[6] Note that the term "map" denotes a particular kind of function in MapReduce programming model.

```
Input: A place name of interest P and a set of textual documents or
       semi-structure files which contain place descriptions
Output: A list of words co-occurrence counting
        [⟨P,wordToken⟩, frequency]
/* The Map procedure */
Mapper(String P, String filename)
List⟨String⟩ T = Tokenize(filename);
forall the wordToken ∈ T do
    forall the eachrow ∈ filename do
        /* Determine whether the words co-occur with P in each record*/
        if (both wordToken and P in row) then
            emit ((String)wordToken, (Integer) 1);
        end
    end
end
/* The Reduce procedure */
Reducer(String wordToken, List⟨Interger⟩ values)
forall the wordToken ∈ T do
    Integer frequency = 0;
    /*Sum all key/values from distributed nodes*/
    forall the value ∈ values do
        frequency = frequency + value;
    end
    emit (String wordToken, Integer frequency);
end
return ⟨P,wordToken⟩, frequency
```

In addition, in order to enable spatial-analysis functions on Hadoop, the Hadoop core is extended to handle geometric features and operations. We choose Esri's open source geometry library because of its popularity in GIS and as a reliable framework in the whole ecosystem (more detailed information in Section 3.2).

(3) Cloudera Manager Web User Interface (CMWebUI): Cloundera Manager[7] is an industry standardized administration package for the Hadoop ecosystem. With CMWebUI, we can deploy and centrally operate the Hadoop infrastructures. In addition, it gives us a cluster-wide, real-time view of nodes and monitors the running services, and enables configuration changes across the cluster. Fig. 2 shows its Web user interface.

---

[7] download at http://www.cloudera.com

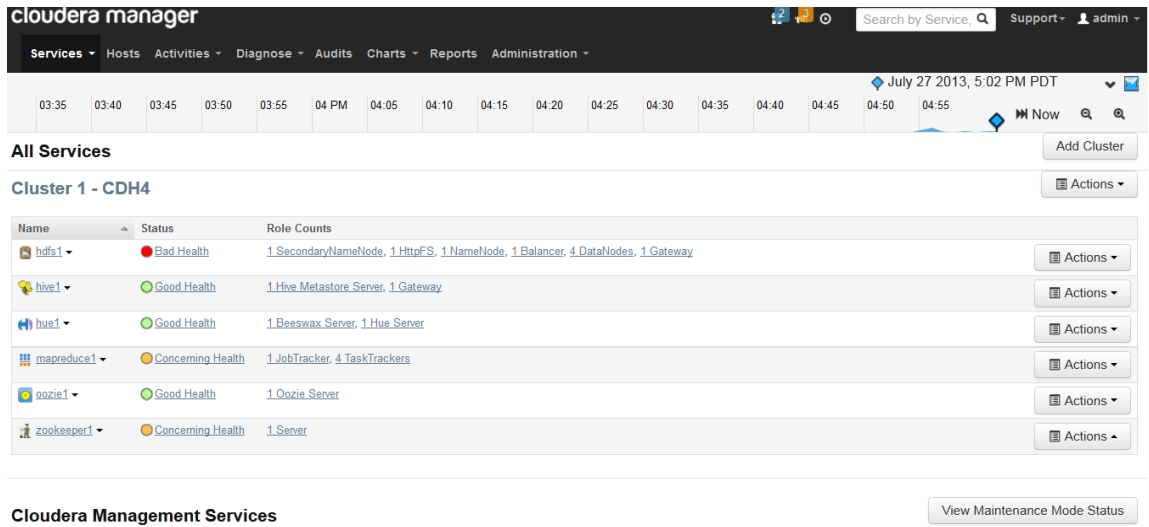

**Fig. 2.** The Cloudera Manager Web user interface

(4) The GIS client supports the geo-visualization of MapReduce operation results transmitted from the Hadoop cluster and built-in geoprocessing models. By enabling HDFS related tools, it also supports converting map features (points, polylines, polygons) into Hadoop-supported data formats for further spatial operations.

*3.2 Enabling spatial analysis on Hadoop*

First, since HDFS cannot directly support the standard GIS data formats, e.g., Esri shapefiles, we need to store the geospatial data in a different way. GeoJSON[8] is an open format for encoding simple geometry features (points, polylines, polygons, and collections of these types) along with their non-spatial attributes. It is an extension of the JavaScript-Object-Notation (JSON) format which is often used for serializing and transmitting structured data over a network connection and meets the HDFS requirements. Both of the spatial and attribute information are stored in plain text as below:

**GeoJSON file examples:**

*{"type": "Feature",*

---

[8] http://www.geojson.org

```
    "geometry": {
      "type": "LineString",
      "coordinates": [[-122.52, 37.71], [-103.23, 41.52], [-95.86, 43.13], ······ ],
    "fields": {
      "prop1": "value",
      "prop2": "string"}
  }
```

Next, we incorporate the GIS tools for Hadoop that have been released on the open-source project site Github[9], which provides an open-source toolkit for Big Spatial Data Analytics powered by Esri and was released in March 2013. We integrate two types of Esri toolkits on Hadoop to handle spatial data: *Geometry API for Java* and *Geoprocessing Tools for Hadoop*. On the server side, the *Geometry API* is a generic library that supports geometry types and basic spatial operations and will allow us to build the MapReduce model for parallel processing of gazetteer entries (including such operations as spatial filter and spatial join). Table 1 lists the spatial relationship analysis and operations that the existing toolkit supports.

**Table 1** The Esri Geometry API supported functions

| Relationship Analysis | equals, disjoint, touches, crosses, within, contains, overlaps |
|---|---|
| Spatial Operations | buffer, clip, convexhull, intersect, union, difference |

The MapReduce algorithm for spatial joins based on the Esri geometry library and the Hadoop system is demonstrated in Algorithm 2. This algorithm is important to analyze the spatial distribution of extracted gazetteer entries and to assign them to the administrative boundaries of places. A spatial join involves matching attribute information from the join feature to the target feature based on their spatial relationships.

---

[9] http://esri.github.io/gis-tools-for-hadoop

The spatial join usually builds on sequentially identifying the spatial relationship between two input features. However, with the help of MapReduce model, this operation can be deployed in the parallel environment. There are two specified functions for the implementation of MapReduce-based spatial join on HDFS:

*The Mapper* function splits the target feature (e.g., a polygon representing a US state) into different keys, i.e. the unique identifier (e.g., the state name). Then, it performs the sub-process of determining whether the target feature contains the join feature, and assigns a key/value (e.g., state name/ counts of points inside). Note not only that the target feature has been split into different keys but also that the join features can be divided into small blocks on HDFS for parallel computation to improve operational efficiency.

*The Reducer* function performs a summary operation (e.g., counting joined point features to each polygon) by aggregating the key/values produced by the *Mapper*.

**Algorithm 2:** MapReduce algorithm of the spatial join operation.

```
/* Example of spatially join points to polygons */
Input: A set of point files PF and a target JSON file of polygon T
Output: A list of [PolygonID, ContainedPointIDs] and the count of
        points in each polygon
polygonFeature=EsriFeatureClass.fromJson(T);
List⟨String⟩ PolygonKey = Tokenize(polygonFeature);
List⟨String⟩ PointID = Tokenize(PF);
/* The Map procedure */
Mapper(Key PolygonKey, String PF)
forall the eachrow ∈ PF do
    Geometry point = new Point(longitude, latitude);
    /*Judge spatial relations*/
    if (GeometryEngine.contains(polygonFeature[PolygonKey], point,
    spatialReference) then
        emit (String PolygonKey, Integer 1, PointID)
    end
end
/* The Reduce procedure */
Reducer(String PolygonKey, List⟨Integer⟩ values, List⟨String⟩
PointIDs )
forall the PolygonKey ∈ T do
    Integer count = 0;
    /*Aggregate all key/values from distributed nodes*/
    forall the value ∈ values do
        count = count + value;
    end
    emit (String PolygonKey, Integer count);
end
/* Optional for appending attributes */
forall the point ∈ PointIDs do
    polygonFeature[PolygonKey].attributes.get(point.attributes)
end
return PolygonKeys,ContainedPointIDs, counts
```

*3.3 A new geoprocessing workflow for Hadoop*

The Hadoop ecosystem lacks a tool to visualize the geospatial footprints of gazetteer entries. An intuitive way is to send the operation results back from the HDFS server to a GIS client such as ArcMap. In addition, the ArcMap software provides hundreds of spatial analysis tools for discovering patterns hidden in the geospatial data. The recently released toolkit *Geoprocessing Tools for Hadoop*[10] established the connection between the ArcGIS environment and the Hadoop system. In our implementation, these tools are used for further analyzing and visualizing the gazetteer entries extracted from the Hadoop system. More importantly, scalable geoprocessing workflows can be created by linking the Hadoop related functions with GIS tools. For example, Fig. 3 presents a geoprocessing workflow running on ArcGIS to submit a MapReduce job for the spatial-join operation (points in polygons) on Hadoop. The main procedures are described as follows:

(1) Features to JSON: Convert the target polygon features from standard ArcGIS format (shapefile) into the GeoJSON format.

(2) Copy Data to HDFS: Transmit the polygon's GeoJSON file based on the WebHDFS mechanism, which uses the standard Hyper-Text Transfer Protocol (HTTP) to support all HDFS user operations including reading files, writing data to files, creating directories, and so on. The user needs permission to access the Hadoop Namenode host server and to operate the HDFS.

(3) Execute Workflow: This tool needs an Oozie[11] Web URL within the Hadoop cluster and an input file that describes the Hadoop Oozie job properties, including the user name, the job-tracker information; and the directories of input features, output features, and the supported library of operations (i.e., the Esri Geometry API for Java package in this case).

(4) Copy Results from HDFS: It transmits the output of aggregating key/value pairs (e.g., counts of points in each polygon) of the MapReduce operation from the Hadoop server to the GIS client.

(5) Join Field: It integrates a GIS function "Join" to append the MapReduce processing results to the target features by matching the key field (e.g., the name of each polygon). As the output of this geoprocessing workflow the aggregated features will be automatically added to display in the ArcGIS environment.

---
[10] https://github.com/Esri/geoprocessing-tools-for-hadoop

---

[11] Oozie job workflow is a collection of actions (i.e. MapReduce jobs, Pig jobs) arranged on Hadoop system and allows one to combine multiple jobs into a logical unit of work.

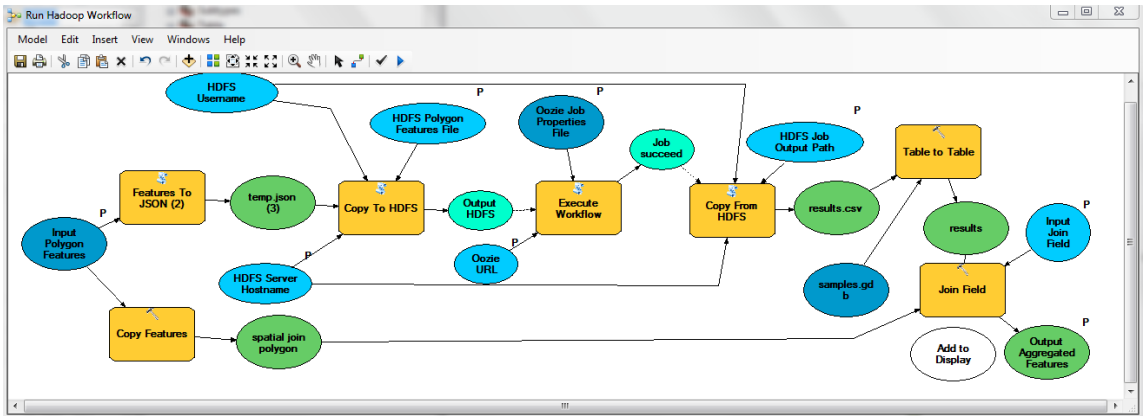

**Fig. 3.** The geoprocessing workflow running on ArcGIS to submit a MapReduce job of the spatial-join operation on Hadoop.

The geoprocessing workflow of spatial join for Hadoop facilitates fast processing and statistics of gazetteer entries. Enabled by this new distributed geoprocessing framework, other computationally intensive spatial analysis tasks can be substantially speeded up, after being decomposed into sub-processes according to the MapReduce paradigm.

## 4. Experiments and Results

In this section we apply the methods introduced above to extract gazetteer entries from the geotagged data in Flickr. First, we extract prominent feature-types using the scalable geoprocessing workflow based on Hadoop. Then, we illustrate how to harvest different geometric types of specified gazetteer entries.

*4.1 Datasets and Hadoop cluster*

A Web crawler was used to collect the geotagged data and store them on HDFS as one type of volunteered gazetteer source. In total, we collected 5,319,623 records within the bounding box of the contiguous US. The photos were either georeferenced by built-in GPS in cameras or manually georeferenced by a user who identified the photo location on the Flickr website. The location could either be the place where a photo was taken or the location of an object in the photo. Automatic recording by a GPS receiver always results in the former case, while manually georeferenced photos could be either way. The Photo metadata includes photo ID, title, description, tags, time when a photo was taken and uploaded, latitude and longitude, as well as lineage information about the users who uploaded the picture (Table 2).

Based on the system architecture introduced above, on the server side, we built a Hadoop cluster by installing, deploying, and configuring the Cloudera Hadoop packages (CDH Version 4.0) on each distributed server and assigning different roles *Namenode, Datandoe, HDFS services, MapReduce services, jobTracker and taskTraker* to them (Table 3). The chief merits of such a Hadoop ecosystem derive from its robustness and scalability at a low cost, by employing multiple normal computer servers instead of a single high-performance cluster. In addition, the system architecture is so flexible that the CDH packages can be deployed either on our local servers in different physical locations or on Amazon EC2 instances as virtual servers.

**Table 2** The metadata structure and an example of Flickr geotagged data

| PhotoID | 5326171618 |
|---|---|
| Title | DSCN41 |
| Description | Santa Barbara Wharf |
| Tags | California, CA, trip, sea, USA, pier, sunset, seafood |
| Taken Time | 12/30/2010 10:39 |
| Uploaded Time | 1/4/2011 20:22 |
| Latitude | 34.4101 |
| Longitude | -119.6856 |
| UserID | 57900412 |

**Table 3** The roles of 10 distributed servers connected on the Hadoop cluster

| Name (count of servers) | Roles | Location | Server Info |
|---|---|---|---|
| UCSBMasterNode (1) | Namenode, HDFS, MapReduce, JobTraker | Santa Barbara | CentOS 5.8, 64 bit, 7.8 GB memory, 3.6 GHz processor, 2 TB storage |
| ASUDataNode (1) | Secondary Namenode, Datanode, HDFS, TaskTraker | Phoenix | CentOS 6.4, 64 bit, 5 GB memory, 2.4 GHz processor, 320 GB storage |
| EC2-RedHat (1) | Datanode, HDFS, TaskTraker | Oregon | CentOS 6.4, 64bit, 7.5 GB memory, 2.4 GHz processor, 420 GB storage |
| EC2-Ubuntu (7) | Datanode, HDFS, TaskTraker | Oregon | Ubuntu 12.04, 64bit, 7.5 GB memory, 2.4 GHz processor, 420 GB storage |

*4.2 Extracting multi-scale spatial distributions of place types*

While authoritative gazetteers provide good quality for long-term administrative place types such as countries, cities, and towns, the crowd-sourced gazetteers could contribute small-scale place types such as restaurants and coffee shops. In order to demonstrate the performance of the new geoprocessing workflow for Hadoop introduced in Section 3.3, we extract and analyze the spatial distribution of some prominent place types (Table 4) in the US, including parks, schools, museums, coffee shops, streets, and rivers. Their frequencies of occurrence are high enough in the tags for a reliable extraction.

After loading the extracted text files of feature types on HDFS according to their keywords (listed in Table 5), we can visualize the geographic footprints of place types and obtain statistical information by running the geoprocessing workflow of spatial joins for Hadoop. The spatial distributions of geotagged points annotated with these feature types in the map extent of the continuous US are shown in Fig. 4. It gives a sense of spatial context for these place types and needs to zoom in the map for exploring more detailed place information in a GIS environment. Named-entity recognition (NER) techniques can be used to further extract place entities. As we know, places are hierarchically organized. Spatial joins can also help to assign the hierarchical names of different geopolitical divisions (such as states, counties, and ZIP code regions) to each gazetteer entry. Table 4 presents a summary of the operational results.

By comparing the computation time of Hadoop-based spatial join operations with that of single *desktop* PC-based spatial join procedures running on a modern laptop with 64-bit operating system, 2.5 GHz Intel-dual-core processors, and 4 GB instant memory, as shown in Fig. 6 (A), we find that the MapReduce-based workflow running on our Hadoop cluster can reduce computing time by an order of magnitude when the number of submitted geotagged points for each place types is sufficiently large (e.g., we saved about 73% of the computing time for 100,000 points). Interestingly the performance of 10 nodes compared with that of 4 nodes on the Hadoop cluster has a comparatively small effect. If we increase the number of target polygons, the Hadoop-based aggregation reduces about half of the time and this is most likely because of the difference in memory (RAM). A specific example of spatially aggregating the 229694 geotagged points of parks to different granularities of US census units -- states (51 polygons), counties (3143 polygons), ZIP code regions (32086 polygons), and census tracts (72851 polygons) -- is shown in Fig. 5. The computation time curves are depicted in Fig. 6. (B). Note that we only connected a relatively small numbers of (four and ten) servers connected to the Hadoop cluster so far, and that higher computation efficiency might be achieved by adding more data nodes equipped with HDFS and task-Trackers. However, Hadoop-based systems often encounter a disk bottleneck in reading data from the network (IO-bound) or in processing data (CPU-bound). An optimized configuration of the Hadoop cluster could improve the cloud computing performance but is not within the scope of this paper; see Kambatla et al. (2009) for more details. Using this example, we demonstrated the high performance of the new scalable geoprocessing workflow based on the MapReduce model and how to derive feature-type-based gazetteer entries inside administrative polygons with GIS tools for Hadoop.

**Table 4** Extracting and analyzing place types from photo tags at different scales

| Feature Types | Keywords | Records | # State | # County | # ZIP |
|---|---|---|---|---|---|
| parks | park, 公园(Chinese), parc (French), parquet (Spanish) | 229694 | 4688 per state 49 states | 145 per county 1580 counties | 33 per ZIP 7042 ZIPs |
| schools | school, university | 112885 | 2304 per state 49 states | 109 per county 1036 counties | 32 per ZIP 3500 ZIPs |
| museums | museum | 65695 | 1341 per state 49 states | 91 per county 722 counties | 39 per ZIP 1706 ZIPs |
| coffee shops | coffee, cafe, coffeehouse, coffeebar, starbucks | 19523 | 398 per state 49 states | 25 per county 788 counties | 7 per ZIP 2643 ZIPs |
| streets | street, road, blvd, freeway, highway | 181410 | 3702 per state 49 states | 92 per county 1980 counties | 6 per ZIP 31941 ZIPs |
| rivers | river, watershed | 45252 | 924 per state 49 states | 37 per county 1217 counties | 14 per ZIP 3371 ZIPs |

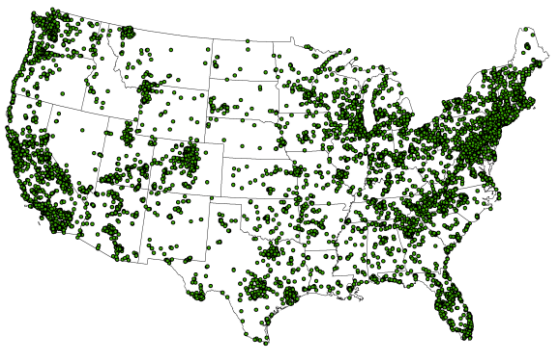 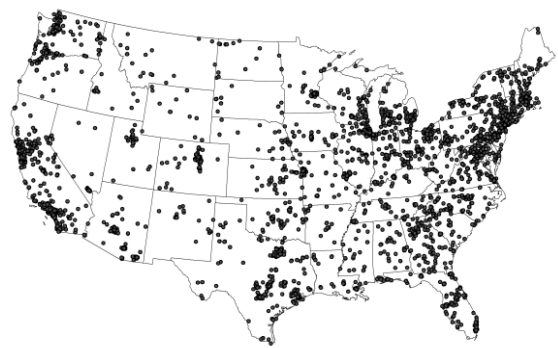

(A)  (B)

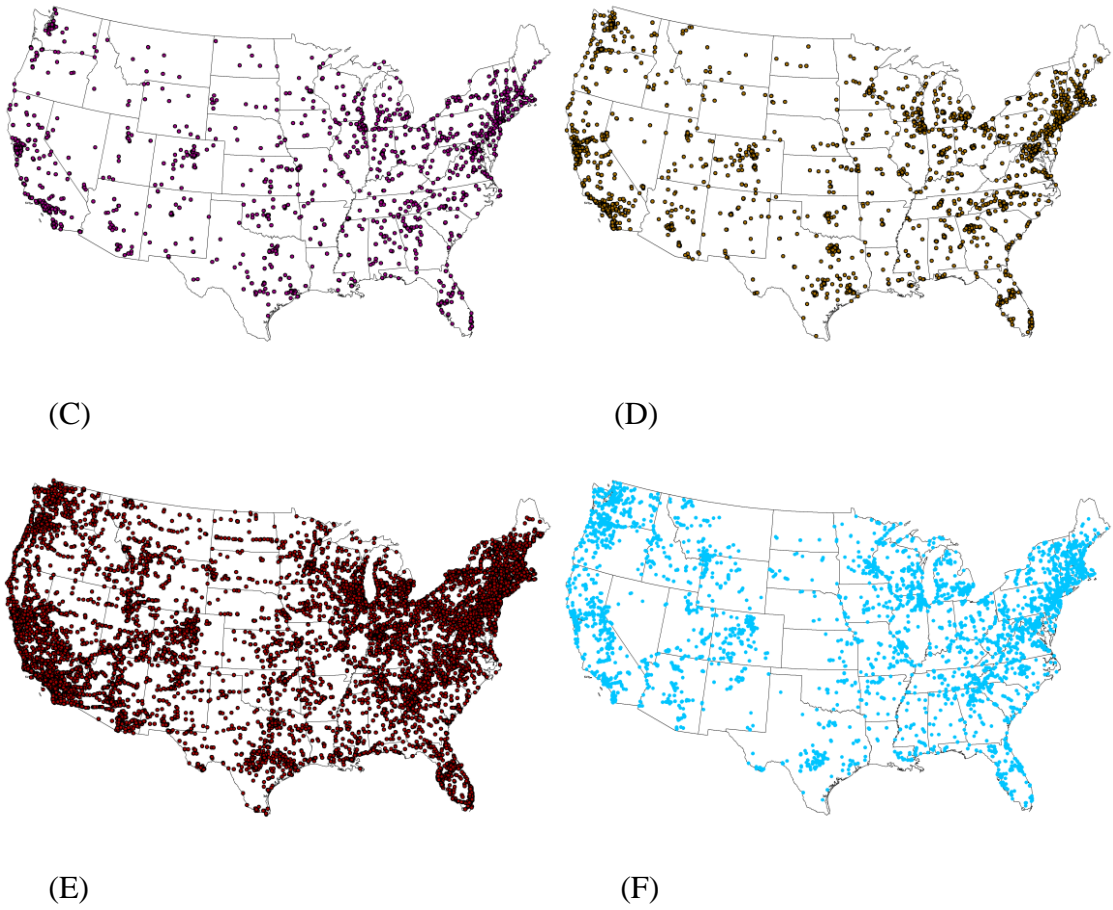

(C)    (D)

(E)    (F)

**Fig. 4**. The spatial distributions of geotagged points annotated with these feature types: (A) parks; (B) schools; (C) museums; (D) coffee shops; (E) streets; (F) rivers.

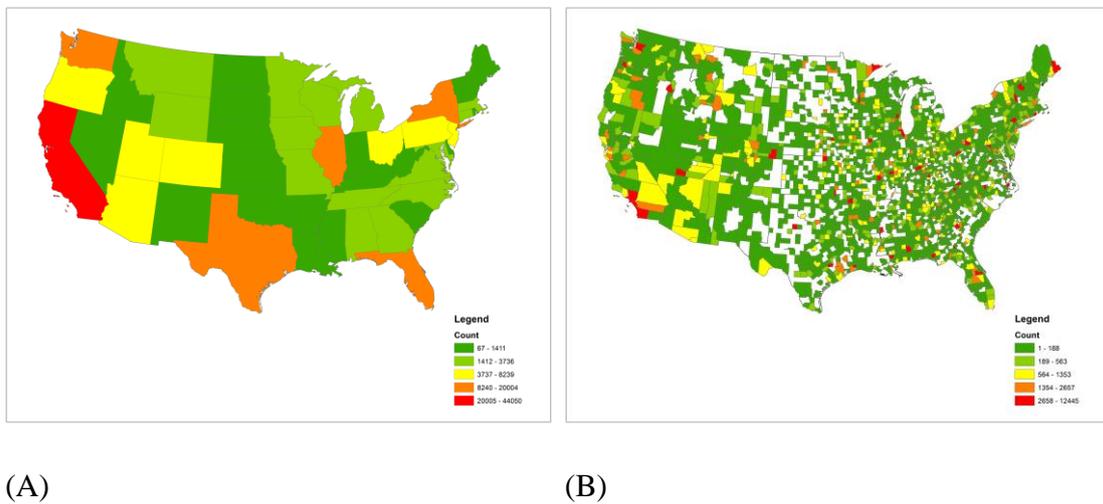

(A)    (B)

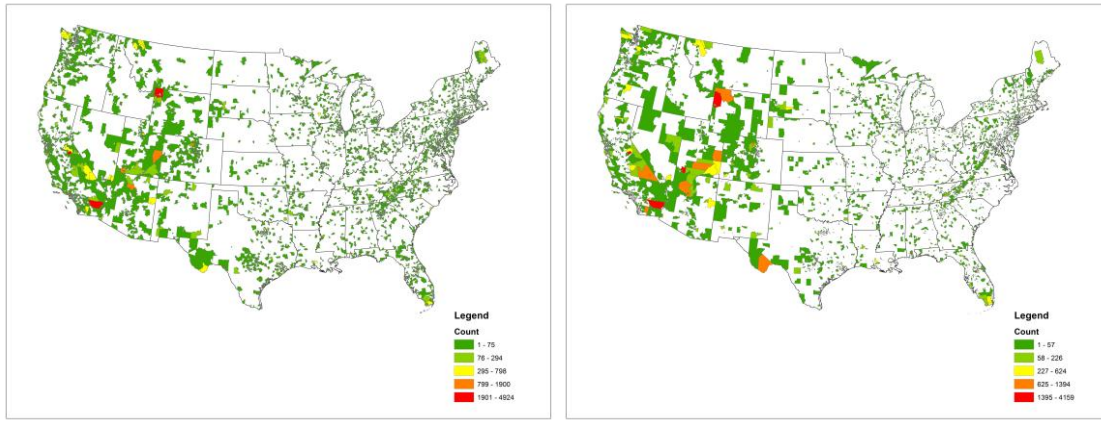

(C)  (D)

**Fig. 5.** The results of spatial join workflow based on Hadoop for *parks*: (A) by US states; (B) by US counties; (C) by US ZIP codes; (D) by US census tracts. (Source: basemaps are provided by Esri)

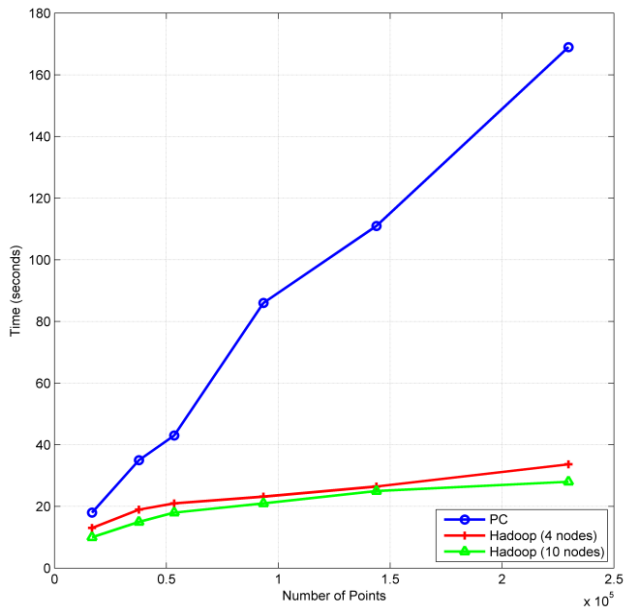 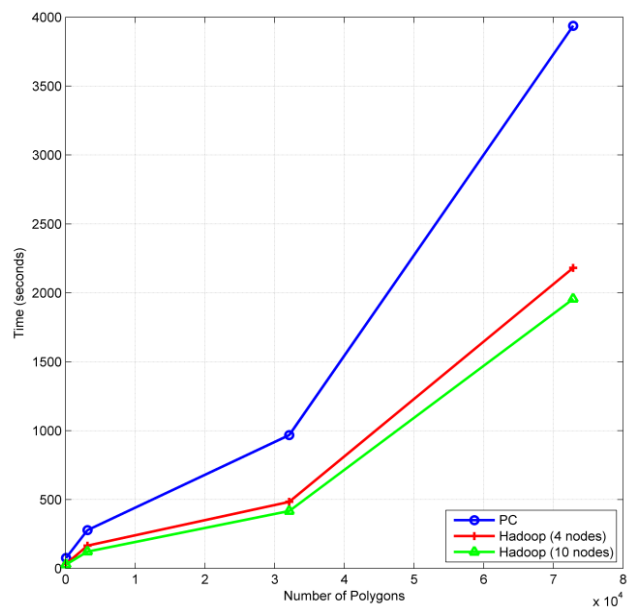

(A)  (B)

**Fig. 6.** The computation time curves of Hadoop-based spatial joins and a single desktop PC: (A) increasing the number of joined points; (B) increasing the number of target polygons.

*4.3 Harvesting gazetteer entries*

The results of place-type-based processing give an overview of the spatial distributions of geotagged points. In order to extract full gazetteer entries, place names, geographic footprints, and feature type descriptions, as well as provenance information are needed. As discussed in Section 2.1, place is a social concept that is perceived and recognized by human beings; therefore, the provenance information about the group of people who identify place is as important as the traditional elements (name, feature type, and footprint). As argued by Goodchild and Li (2012a), the current representation of place entries in a gazetteer independent of the users should be complemented by another element of source. It helps reveal the *binary* relationship between a place and its contributors, i.e., to know not only where a place is and how it is referred-to, but also who refers to it in this way. The provenance of gazetteer entries would enhance research on social perception of places because the same (or similar) location may be named differently by different groups of people instead of the traditional *unary* form that only links the place and its official name.

In the following, we illustrate the construction processes for retrieving different geometric (point, polyline, polygon) gazetteer entries annotated with Santa Barbara Courthouse, California State Route 1 (SR1 or Highway1), and Harvard University. Table 5 presents the summary of harvested crowd-sourced gazetteer entries with the given keywords. The geographic footprints and place descriptions were extracted from the GPS locations and the tags that were given to a place. The provenance information was derived from the users who contributed the geotagged photos to a given place. The collected provenance information from users will help to further validate extracted entries based on quality assurance methods as well as trust model (more details are provided in Section 4.4).

**Table 5** The harvested different geometry types (point, polyline, polygon) of crowd-sourced gazetteer entries

| Place names | Geographic footprints | Place descriptions (top 10 ranked tags) | Provenances (only list the number |

| | | | of contributors here) |
|---|---|---|---|
| Santa Barbara Courthouse | {Point:[GeoJSON]} | Santa Barbara courthouse California county palm trees view historical architecture | 81 points from 22 trusted UserIDs |
| California State Route 1 | {Line: [GeoJSON]} | highway1 California Sanfrancisco bigsur motorcycleride hearstcastle beach ocean coast USA | 427 points 59 trusted UserIDs |
| Harvard University | {Polygon:[GeoJSON] } | Harvard University Cambridge USA Boston Massachusetts Square Harvard-Westlake Flintridge Sacred | 637 points from 176 trusted UserIDs |

Santa Barbara Courthouse, located at downtown Santa Barbara, is a local historic landmark and famous for its architecture and the panoramic view of the city. It is better to take it as a point gazetteer entry although multiple geotagged-photo points are extracted and most of them distributed around the main building (Fig. 7). We applied the *Standard Deviational Ellipse* (SDE) statistical analysis to identify the significant points, which is more robust to outliers and could summarize the central tendency and directional trend of point distributions (Mitchell, 2005). Next, we selected the points (SPs) contained by the two standard deviation (2σ-SDE) polygon which covers approximately 95 percent of the extracted points. Finally, a 2σ-centroid of SPs in the identified cluster was assigned to the geographic footprint for this feature. In addition, by counting the frequency of tags, we perceive that location-context words (Santa Barbara, California, county), local distinguishing features (palm trees) and the characteristics of the landmark itself (view, historical, architecture) are the most frequently used texts to express the users' feelings and experiences about a place.

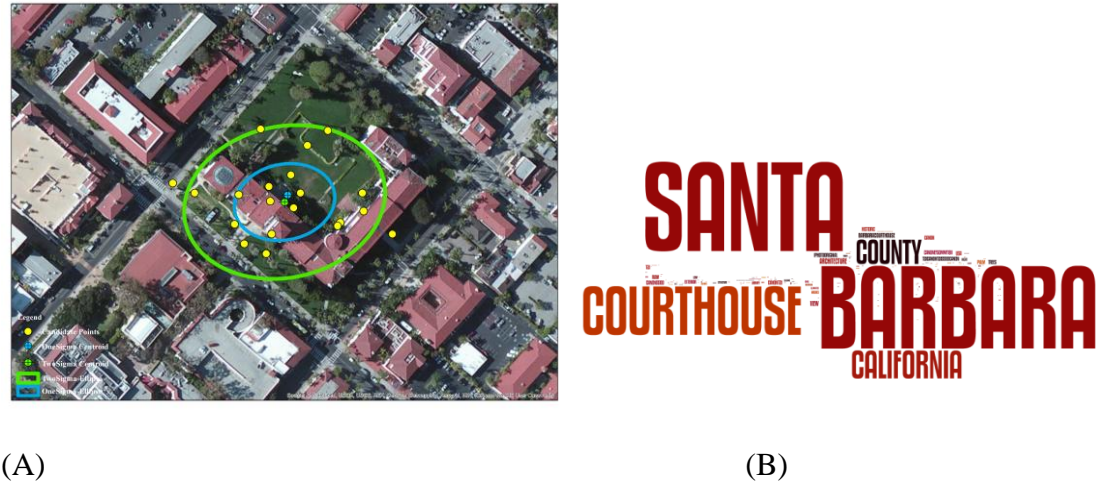

(A)                                   (B)

**Fig. 7.** The geographic footprint and tag descriptions for *Santa Barbara Courthouse*: (A) extracted geotagged points for this feature and its 1σ-centroid (Blue) and 2σ-centroid (Green) with the standard deviational ellipses; (B) a word-cloud visualization of the extracted tags using the Wordle[12] tool.

California SR1 is one of the most famous highways along the Pacific Coast in the US. By merging the geotagged points labelled 'highway1' or 'freeway1' and filtering them by the geographic footprint of California, the automatically generated line presents a good shape of the main SR1 (Fig. 8). A denser spatial and temporal sampling of geotagged points and more strict algorithms may provide a better and more complete footprint of the route. More importantly, by exploring the semantic tags, we can derive fruitful feature attributes and social descriptions for fast updating of road gazetteer entries. For SR1, we get the information about where the entry is located (USA, California), the main cities (San Francisco, Los Angeles) and famous landmarks (Big Sur, Hearst Castle) along the route, as well as other descriptive characteristics (motorcycle ride, beach, ocean, coast). This process is unlike traditional automatic road updating techniques with GPS trajectories (Cao & Krumm, 2009) which only contain the geometry information.

---

[12] http://www.wordle.net

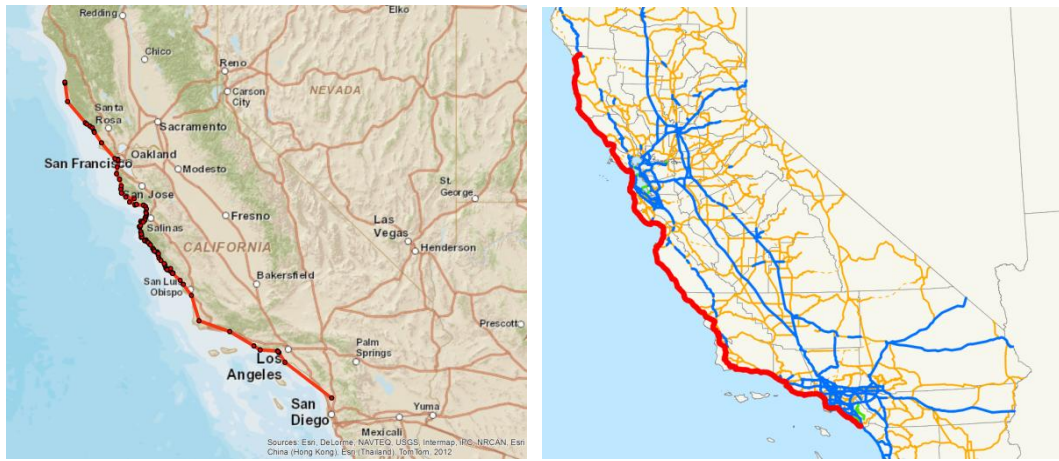

(A)                                    (B)

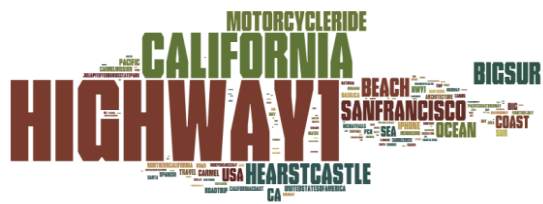

(C)

**Fig. 8.** The geographic footprint and tag descriptions for *California SR1*: (A) the automatically constructed line feature by connecting all points following the longitude sequence; (B) the California SR1 map from Wikipedia; (C) a word-cloud visualization of the extracted tags using the Wordle tool.

The final example is Harvard University. In crowd-sourced gazetteers, in order to store the more complete extent of the university campus, it should be represented as a polygon. As shown in Fig. 9 (A), the extracted geotagged points labeled with 'Harvard University' are distributed among the central campus, on Harvard Bridge and along other scattered locations. Several methods have been proposed to generate the polygonal representation of places from footprint points. For example, kernel-density estimation has been introduced (e.g., Jones et al., 2008; Li & Goodchild, 2012) to extract the boundaries of vague places according to a threshold point density. Keßler et al. (2009) assigned centroid locations to geotags and used Delaunay triangulation graph to identify clusters in the point clouds. Liu et al. (2010) proposed a point-set-based-region model to approximate vague area objects.

Here, we introduce a fuzzy-set-based method to extract geographic footprints of polygonal places. Fuzzy-set-based classification and identification methods have been widely used in GIS and related disciplines (Burrough & Frank, 1996; Cross & Firat, 2000; Robinson, 2003; Montello et al., 2003). The fuzzy set *A* can be interpreted as the degree of membership of *X* in a set; values assigned fall within the range [0, 1]. Many membership functions to express the grade of membership of *X* in a fuzzy set *A* have been discussed by Robinson (2003). For the crowd-sourced gazetteer entries, the geotags of a place generated by users usually follow a clustering structure, thus we suggest using a distance-decay function (Taylor, 1971; Leung & Yan, 1997) to measure the membership of candidate point locations assigned to a place:

$$\mu(x) = \begin{cases} 1, (0 \leq d_x \leq d_1) \\ \dfrac{C}{d_x^{\beta}}, (d_1 < d_x < d_2) \\ 0, (d_x \geq d_2) \end{cases}$$

where $d_x$ is the distance between a candidate point and the centroid point of the cluster, $\beta$ is a decay parameter, and *C* is a parameter to scale the range of membership scores. We need to set distance thresholds $d_1$ and $d_2$.

To store the spatial footprint of a polygonal gazetteer entry, we can use the *α*-cut technique (Robinson, 2003). A crisp set $A_\alpha$ contains all elements of *X* whose membership scores in $A_\alpha$ are greater than or equal to *α*. The α-cut-boundary of a place can be further derived from the points in $A_\alpha$ based on the minimum-enclosing-geometries, such as the *α*-cut-minimum-bounding-rectangle, or the *α*-cut-convex hull. Here, we set *β*=1, $d_1$=50 meters, $d_2$=5000 meters, and *C*=5 (note that the parameters might vary at different scales). Fig. 9 (B) and (C) present two different shapes of α-cut-boundaries: the *α*-cut-minimum-bounding-rectangle and the *α*-cut-convex-hull. All the 0.5-cut-boundaries have a good representation of the footprint of the northern Harvard campus (not including the southern part separated by the Charles River), while the 0.8-cut-boundaries indicate the core attractive areas where the geotagged photos are taken.

After updating the geographic footprint, we also need to capture the users' descriptions about Harvard University. Besides conventional place descriptions that are related to place names and local landmark characteristics introduced above, the comments with tags related to events can also be detected. For example, during the temporal extent of downloaded data, there was a girls' basketball match between the Flintridge-Sacred-Heart team and the Harvard-Westlake team hosted at Harvard on January 21, 2011. Consequently, Flickr users uploaded many geotagged photos with comments and place descriptions about this particular match. This is why we get a high frequency of tags: Flintridge-Sacred-Heart and Harvard-Westlake at Harvard.

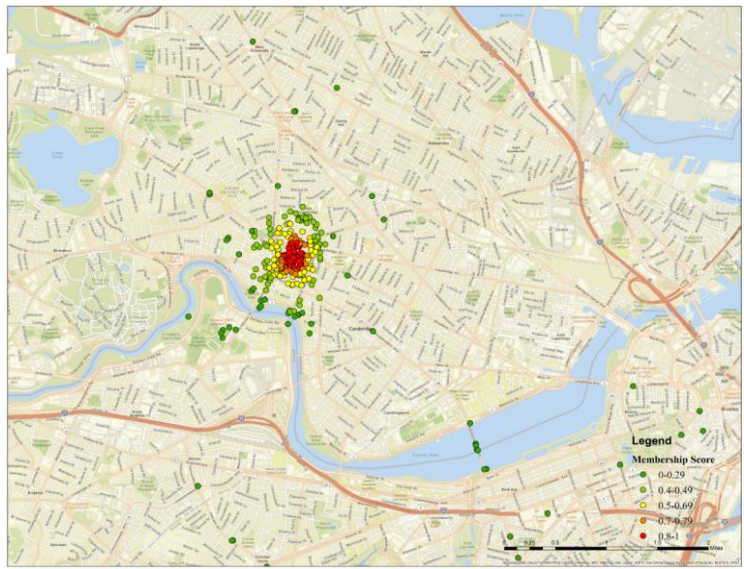

(A)

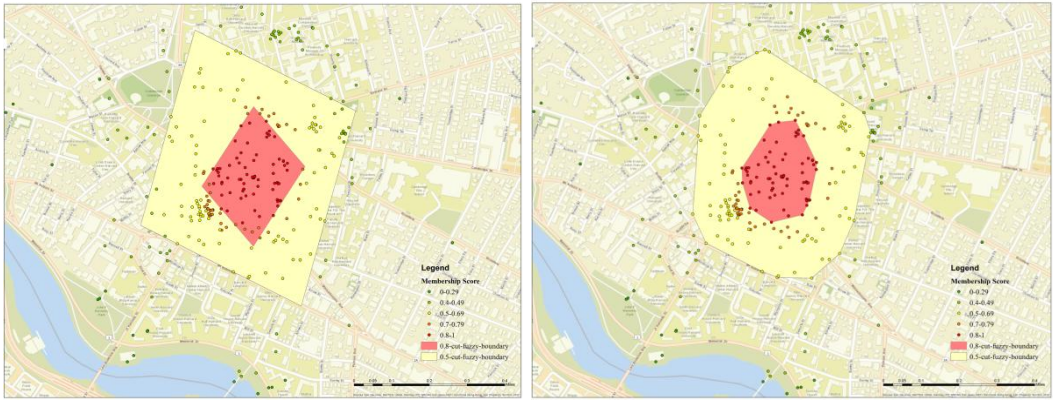

(B)                              (C)

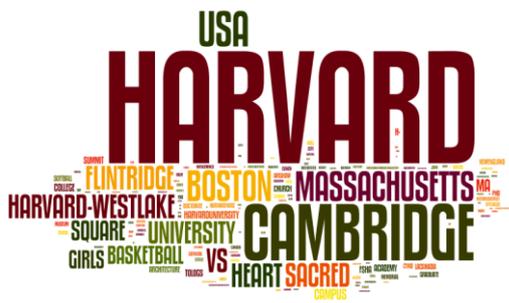

(D)

**Fig. 9.** The geographic footprints for *Harvard University*: (A) the spatial distribution of geotagged points with their fuzzy membership scores; the 0.5 and 0.8-cut-boundaries represented by (B) the minimum bounding rectangle; (C) the convex hull; (D) A word-cloud visualization of the extracted tags using Wordle tool. (Note that different projections between basemap and minimum bounding geometries make their shapes become deformed.)

*4.4 Outlook on the provenance-based trust evaluation*

VGI as a data source preserves the semantic diversity in the contributors' cognition of places. The data are created through a large volume of voluntary contributions and quality issue has been widely discussed by the VGI research community. Goodchild and Li (2012b), for instance, discussed three approaches for the quality assurance: *crowd-sourcing*, *social*, and *geographic* methods. In the absence of ground-truth data, several studies have proposed the use of provenance information to estimate the quality of VGI. For example, researchers suggested using contributor-associated trust to measure crowd-sourced data quality. Mooney and Corcoran (2012) investigated the tagging and annotation of OSM features using provenance. Keßler and Groot (2013) proposed a five-indicator trustworthiness model as a proxy in the case study of OSM. The results of an empirical study support the hypothesis that VGI data quality can be assessed by using a trust model based on the provenance information.

In this work, we have collected the provenance metadata for each gazetteer entry, i.e., the contributors, the total number of uploaded photos and time-stamps of contributions. Like other crowdsourcing platforms, a small number of "active users" share most contributions which follow a power-law distribution ranked by the number of uploaded photos (see Fig. 10); only 8% of the total 440000 contributors have shared more than 10 geotagged photos in the collected datasets.

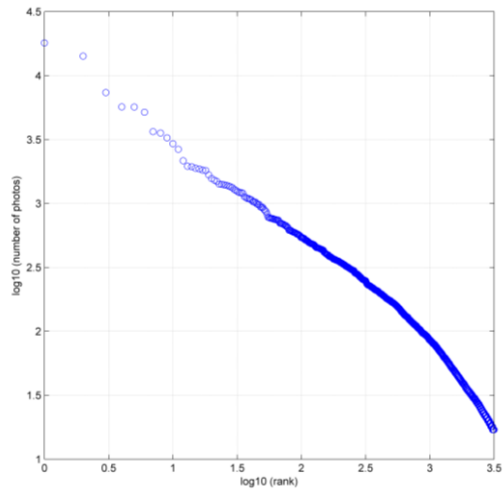

**Fig. 10.** The power-law distribution of generated photos by top-ranked users (on log-log plot)

In contrast to OSM or Wikipedia, the contributors' reputation and trustworthiness cannot be assessed by revisions; in Flickr, we can only rely on the contributors' past geotagging and photo sharing behaviors to establish a user-reputation model: a user $i$ have reputation value $R_i(t)$ at time t.

$$R_i(t) = \frac{\text{the number of reliable geotagged photos}(N_{ir})}{\text{the total amount of photos which a user has uploaded}(N_i)} * W_{rank}$$

A reliable geotagged photo means that its position accuracy meets the quality criteria and consists with the geographic knowledge (Goodchild & Li, 2012b). $W_{rank}$ is a weighted rank based on total contribution; the *active users* who contribute more photos have higher value of $W_{rank}$. We trust the content generated by high reputation users for crowd-sourced gazetteer construction and enrichment. In addition, for each gazetteer entry, we set up a bottom-line requirement: with minimum number (15) of contributors and a minimum number (10) of tag descriptions according to the observation of overall characteristics in the sample datasets (Table 5). Further filtering work and recalculation will be processed based on the contributors reputation scores. We presented an intuitive way to filter reliable geotagged content. Alternative, more complex trust models based on the provenance metadata will be addressed in our future work.

**5. Conclusions and Future work**

In summary, space and place are associated through gazetteers in a wide variety of geospatial applications. While traditional gazetteers that are constructed and maintained by official authorities lack informal and vernacular places, we demonstrate a Big Data-driven approach by mining VGI sources to create a crowd-sourced gazetteer. Three examples of different types (point, polyline, polygon) of geographic features are extracted, analyzed and visualized in this study. We also present an intuitive user reputation model for the trust evaluation.

This semi-automatic construction of a crowd-sourced gazetteer can be facilitated by using high-performance computing resources because it involves the process of mining large-volumes of geospatial data. We designed and established a Hadoop-based processing platform (GPHadoop) to show the promise of using VGI and cloud computing in gazetteer research and GIScience in general. In particular, our approach has the following merits:

- Using the examples of the spatial join operation to the increasing number of points in different geographic scales, we demonstrate that the MapReduce-based algorithm has a higher efficiency to process such Big Geo-Data analysis compared to a traditional desktop PC-based analysis.
- The MapReduce algorithm of counting co-occurrence words makes it possible to rapidly extract parts of a place semantics and popular tags to characterize a place.
- The platform enables scalable geoproccessing workflows to solve geospatial problems based on the Hadoop ecosystem and Esri GIS tools, which make contributions in connecting GIS to a cloud computing environment for the next frontier of Big Geo-Data Analytics.

There are four major areas that require further work: (1) the conflation and integration of crowd-sourced gazetteers that include more place entries and fruitful descriptions extracted from various sources, (2) the exploration of other spatial analysis functions that can be executed on Hadoop, (3) gazetteer schema (ontologies) that go beyond names, footprints, and types, and (4) research about efficiency and quality assurance issues. In this research, only two MapReduce algorithms and 10 connected-server-nodes were implemented on the Hadoop cluster for processing Flickr geotagged data; further research is required to explore which types of operations are appropriate to such parallel computing systems for Big Geo-Data analysis and what the performance of Hadoop cluster is if increasing to hundreds of nodes, as well as to incorporate more heterogeneous volunteered data sources for constructing more holistic perspectives on places.

**Acknowledgements**
The authors would like to thank Michael F. Goodchild for discussions, comments and guidance on this research; thank Shaowen Wang for the introduction of Cloudera

Hadoop system into our work; and also thank Michael Park, David Kaiser, Mausour Raad and Marwa Mabrouk for their great help in the implementation of Esri GIS tools for Hadoop. Finally, comments and suggestions from the editors, reviewers, and our colleagues at UCSB's STKO lab helped us to reorganize and make improvements to the manuscript.

**References:**

Adams, B. and McKenzie, G. (2013). Inferring thematic places from spatially referenced natural language descriptions. In *Crowdsourcing Geographic Knowledge*, pp. 201-221. Springer Press.

Adams, B., & McKenzie, G. (2012). Frankenplace: An Application for Similarity-Based Place Search. In *ICWSM,* AAAI Press.

Adams, B. and Janowicz, K. (2012). On the Geo-Indicativeness of Non-Georeferenced Text. In *ICWSM 2012*, pp. 375-378, AAAI Press.

Agnew, J. (2011). Space and place. In J. Agnew, & D. Livingstone (Eds.), *The SAGE handbook of geographical knowledge* (pp. 316-330). Thousand Oaks: SAGE, (Chapter 23).

Aji, A., Wang, F., Vo, H., Lee, R., Liu, Q., Zhang, X., & Saltz, J. (2013). Hadoop-GIS: A high performance spatial data warehousing system over MapReduce. *Proceedings of the VLDB Endowment*, 6 (11), 1009-1020.

Armbrust, M., Fox, A., Grith, R., Joseph, A. D., Katz, R., Konwinski, A., Lee, G., Patterson, D., Rabkin, A., Stoica, I., & Zaharia, M. (2010). A view of cloud computing. *Communications of the ACM*, 53, 50-58.

Batty, M. (2008). The size, scale, and shape of cities. *Science*, 319, 769-771.

Bernad, J., Bobed, C., Mena, E., & Ilarri, S. (2013). A formalization for semantic location granules. *International Journal of Geographical Information Science*, 27, 1090-1108.

Burrough P.A. & Frank A.U. (Eds.) (1996). *Geographic Objects with Indeterminate Boundaries*. London: Taylor & Francis.

Buyya, R., Yeo, C. S., Venugopal, S., Broberg, J., & Brandic, I. (2009). Cloud computing and emerging it platforms: Vision, hype, and reality for delivering computing as the 5th utility. *Future Generation Computer Systems*, 25, 599-616.

Cao, L., & Krumm, J. (2009). From GPS traces to a routable road map. In *Proceedings of the 17th ACM SIGSPATIAL International Conference on Advances in Geographic Information Systems* (pp. 3-12). ACM.

Cross, V., & Firat, A. (2000). Fuzzy objects for geographical information systems. *Fuzzy Sets and Systems*, 113, 19-36.

Dean, J., & Ghemawat, S. (2008). Mapreduce: simplified data processing onlarge clusters. *Communications of the ACM*, 51, 107-113.

Gao, S., Yu, H., Gao, Y., & Sun, Y. (2010). A design of RESTful style digital gazetteer service in cloud computing environment. In *Proceedings of the 18th International Conference on Geoinformatics* (pp. 1-6). IEEE.

Gao, S., Janowicz, K., McKenzie, G., & Li, L. (2013). Towards platial joins and buffers in place-based GIS. In *Proceedings of the 1st ACM SIGSPATIAL International Workshop on Computational Models of Place (COMP' 2013)* (pp. 1-8). ACM.

Goldberg, D. W., Wilson, J. P., & Knoblock, C. A. (2009). Extracting geographic features from the internet to automatically build detailed regional gazetteers. *International Journal of Geographical Information Science*, 23, 93-128.

Goodchild, M. F. (2004). The alexandria digital library: review, assessment, and prospects. *D-Lib Magazine*, 10.

Goodchild, M. F. (2007). Citizens as sensors: the world of volunteered geography. *GeoJournal*, 69, 211-221.

Goodchild, M. F. (2011). Formalizing place in geographic information systems. In L.M. Burton, S.P. Kemp, M.-C. Leung, S.A. Matthews, and D.T. Takeuchi (Eds), *Communities, Neighborhoods, and Health: Expanding the Boundaries of Place* (pp. 21–35). New York: Springer.

Goodchild, M. F., & Hill, L. L. (2008). Introduction to digital gazetteer research. *International Journal of Geographical Information Science*, 22, 1039-1044.

Goodchild, M. F., & Janelle, D. G. (Eds.) (2004). *Spatially Integrated Social Science.* New York: Oxford University Press.

Goodchild, M.F. & Li, L. (2012a). Formalizing space and place. In P. Beckouche, C. Grasland, F. Guérin-Pace, and J.-Y. Moisseron, editors, *Fonder les Sciences du Territoire*, pp. 83–94. Paris: Éditions Karthala.

Goodchild, M.F. & Li, L., (2012b). Assuring the quality of volunteered geographic information, *Spatial Statistics*, 1: 110–120.

Guo, Q., Liu, Y., & Wieczorek, J. (2008). Georeferencing locality descriptions and computing associated uncertainty using a probabilistic approach. *International Journal of Geographical Information Science*, 22, 1067-1090.

Harrison, S., & Dourish, P. (1996). Re-place-ing space: the roles of place and space in collaborative systems. In *Proceedings of the 1996 ACM conference on Computer supported cooperative work* (pp. 67-76). ACM.

Harvey, F., Kuhn, W., Pundt, H., Bishr, Y., & Riedemann, C. (1999). Semantic interoperability: A central issue for sharing geographic information. *The Annals of Regional Science*, *33*(2), 213-232.

Hastings, J. T. (2008). Automated conflation of digital gazetteer data. *International Journal of Geographical Information Science*, 22, 1109-1127.

Hill, L., Frew, J., & Zheng, Q. (1999). Geographic names: The implementation of a gazetteer in a georeferenced digital library, *D-Lib Magazine*, January.

Hill, L. L. (2000). Core elements of digital gazetteers: placenames, categories, and footprints. In *Research and advanced technology for digital libraries* (pp.280-290). Springer.

Hill, L. L. (2006). *Georeferencing: The geographic associations of information*. Cambridge & London: The MIT Press.

Hubbard, P, Kitchin, R., & Valentine, G. (Eds.) (2004). *Key Thinkers on Space and Place*. London: SAGE.

Janowicz, K. (2009). The role of place for the spatial referencing of heritage data. In *The Cultural Heritage of Historic European Cities and Public Participatory GIS Workshop*, pp. 17-18.

Janowicz, K., & Keßler, C. (2008). The role of ontology in improving gazetteer interaction. *International Journal of Geographical Information Science*, 22, 1129-1157.

Janowicz, K., Scheider, S., Pehle, T., & Hart, G. (2012). Geospatial semantics and linked spatiotemporal data-past, present, and future. *Semantic Web*, 3, 321-332.

Jones, C. B., Alani, H., & Tudhope, D. (2001). Geographical information retrieval with ontologies of place. In *Spatial information theory* (pp. 322-335). Springer.

Jones, C. B., Purves, R. S., Clough, P. D., & Joho, H. (2008). Modelling vague places with knowledge from the web. *International Journal of Geographical Information Science*, 22, 1045-1065.

Kambatla, K., Pathak, A., & Pucha, H. (2009). Towards optimizing Hadoop provisioning in the cloud. In *Proc. of the First Workshop on Hot Topics in Cloud Computing* (pp.1-5).

Keßler, C., Janowicz, K., & Bishr, M. (2009). An agenda for the next generation gazetteer: Geographic information contribution and retrieval. *In Proceedings of the 17th ACM SIGSPATIAL international conference on advances in Geographic Information Systems* (pp. 91-100). ACM.

Keßler, C., Maué, P., Heuer, J. T., & Bartoschek, T. (2009). Bottom-up gazetteers: Learning from the implicit semantics of geotags. In *GeoSpatial semantics* (pp. 83-102). Springer.

Keßler, C., & de Groot, R. T. A. (2013). Trust as a Proxy Measure for the Quality of Volunteered Geographic Information in the Case of OpenStreetMap. In *Geographic Information Science at the Heart of Europe* (pp. 21-37). Springer International Publishing.

Leung, Y., & Yan, J. (1997). Point-in-polygon analysis under certainty and uncertainty. *GeoInformatica*, 1, 93-114.

Li, L., Goodchild, M., & Xu, B. (2013). Spatial, temporal, and socioeconomic patterns in the use of Twitter and Flickr. *Cartography and Geographic Information Science*, 40, 61-77.

Li, L., & Goodchild, M. F. (2012). Constructing places from spatial footprints. In *Proceedings of the 1st ACM SIGSPATIAL International Workshop on Crowdsourced and Volunteered Geographic Information* (pp. 15-21). ACM.

Li, W., Goodchild, M. F., Anselin, L., & Weber, K. (2013)
Li, W., Goodchild, M. F., Anselin, L., & Weber, K. (2013). A service-oriented smart cybergis framework for data-intensive geospatial problems. In M. F. Goodchild, & S.Wang (Eds.), *CyberGIS: Fostering a New Wave of Geospatial Discovery and Innovation*. Springer (forthcoming).

Li, W., Raskin, R., & Goodchild, M. F. (2012). Semantic similarity measurement based on knowledge mining: an artificial neural net approach. *International Journal of Geographical Information Science, 26*, 1415-1435.

Li, W., Yang, P., & Zhou, B. (2008). Internet-based spatial information retrieval. *Encyclopedia of GIS*. pp. 596-599. Springer.

Liu, Y., Guo, Q., Wieczorek, J., & Goodchild, M. F. (2009a). Positioning localities based on spatial assertions. *International Journal of Geographical Information Science*, 23, 1471-1501.

Liu, Y., Li, R., Chen, K., Yuan, Y., Huang, L., & Yu, H. (2009b). KIDGS: A geographical knowledge-informed digital gazetteer service. In *17th International Conference on Geoinformatics* (pp. 1-6). IEEE.

Liu, Y., Yuan, Y., Xiao, D., Zhang, Y., & Hu, J. (2010). A point-set-based approximation for areal objects: A case study of representing localities. *Computers, Environment and Urban Systems*, 34(1), 28-39.

Liu, Y. & Wang, S. (2013). A scalable parallel genetic algorithm for the generalized assignment problem. *Parallel Computing*. (in press)

Madden, S. (2012). From databases to big data. *Internet Computing, IEEE*, 16, 4-6.

Mitchell, A. (2005). *The ESRI Guide to GIS analysis, Volume 2: Spartial measurements and statistics*. Redlands: ESRI Press.

Montello, D. R., Goodchild, M. F., Gottsegen, J., & Fohl, P. (2003). Where's downtown?: Behavioral methods for determining referents of vague spatial queries. *Spatial Cognition & Computation*, 3, 185-204.

Mooney, P., & Corcoran, P. (2012). The annotation process in OpenStreetMap. *Transactions in GIS*, 16(4), 561-579.

Ostermann, S., Iosup, A., Yigitbasi, N., Prodan, R., Fahringer, T., & Epema, D. (2010). A performance analysis of EC2 cloud computing services for scientific computing. In *Cloud Computing* (pp. 115-131). Springer.

Pavlo, A., Paulson, E., Rasin, A., Abadi, D. J., DeWitt, D. J., Madden, S., & Stonebraker, M. (2009). A comparison of approaches to large-scale data analysis. In *Proceedings of the 2009 ACM SIGMOD International Conference on Management of data* (pp. 165-178). ACM.

Rey, S. J., Anselin, L., Pahle, R., Kang, X., and Stephens, P. (2013). Parallel optimal choropleth map classification in PySAL. *International Journal of Geographical Information Science*, 27(5), 1023-1039.

Robinson, V. B. (2003). A perspective on the fundamentals of fuzzy sets and their use in geographic information systems. *Transactions in GIS*, 7, 3-30.

Saalfeld, A. (1988). Conflation: automated map compilation. *International Journal of Geographical Information Systems* 2(3): 217–228.

Scheider, S. (2012). *Grounding geographic information in perceptual operations* (Vol. 244). Frontiers in Artificial Intelligence and Applications Series, IOS Press.

Shvachko, K., Kuang, H., Radia, S., & Chansler, R. (2010). The Hadoop distributed file system. In *26th Symposium on Mass Storage Systems and Technologies (MSST 2010)* (pp. 1-10). IEEE.

Taylor, P. J. (1971). Distance transformation and distance decay functions. *Geographical Analysis*, 3, 221-238.

Tuan, Y.-F. (1977). *Space and place: The perspective of experience*. Minneapolis:

University of Minnesota Press.

Uryupina, O. (2003). Semi-supervised learning of geographical gazetteers from the internet. In *Proceedings of the HLT-NAACL 2003 workshop on Analysis of geographic references* (pp. 18-25). Association for Computational Linguistics.

Vasardani, M., Winter, S., & Richter, K.-F. (2013). Locating place names from place descriptions. *International Journal of Geographical Information Science*, DOI:10.1080/13658816.2013.785550.

Wang, S. (2010). A CyberGIS framework for the synthesis of cyberinfrastructure, GIS, and spatial analysis. *Annals of the Association of American Geographers*, 100, 535-557.

Wang, S., Anselin, L., Bhaduri, B., Crosby, C., Goodchild, M. F., Liu, Y., & Nyerges, T. L. (2013). CyberGIS software: a synthetic review and integration roadmap. *International Journal of Geographical Information Science*, 27, 2122-2145.

White, T. (2012). *Hadoop: the definitive guide (Third Edition)*. O'Reilly.

Yang, C., Raskin, R., Goodchild, M., & Gahegan, M. (2010). Geospatial cyberinfrastructure: past, present and future. *Computers, Environment and Urban Systems*, 34, 264-277

Yang, C., Goodchild, M., Huang, Q., Nebert, D., Raskin, R., Xu, Y., Bambacus, M., & Fay, D. (2011a). Spatial cloud computing: how can the geospatial sciences use and help shape cloud computing? *International Journal of Digital Earth, 4*, 305-329.

Yang, C., Wu, H., Huang, Q., Li, Z., & Li, J. (2011b). Using spatial principles to optimize distributed computing for enabling the physical science discoveries. *Proceedings of the National Academy of Sciences*, 108(14), 5498-5503.

Yao, X., & Thill, J.-C. (2006). Spatial queries with qualitative locations in spatial information systems. Computers, environment and urban systems, 30, 485-502.